\documentclass{PoS}

\title{The Crab optical and ultraviolet polarimetry}

\ShortTitle{The Crab optical/UV polarimetry}

\author{\speaker{Roberto P. Mignani}\\
        Mullard Space Science Laboratory, University College London \\
        E-mail: \email{rm2@mssl.ucl.ac.uk}}

\abstract{Polarisation  measurements  of pulsars  and  of their pulsar wind  nebulae (PWNe) are uniquely able to  provide deep insights into the highly magnetised  relativistic environment of young, rotation-powered isolated neutron  stars (INSs).   Besides the  radio  band,  optical observations  are primarily suited to providing such  insights. The first INS for which optical polarisation observations were performed is the Crab pulsar which is also the brightest one ($V=16.5$). For this reason, the Crab pulsar is also the only INS for which repeated,  phase-resolved polarisation measurements have been performed through the years. Moreover, it is the only case, together with the much fainter and distant PSR B0540$-$69 in the Large Magellanic Cloud (LMC), of an optical pulsar embedded in an optical PWN. Thus, the Crab is a perfect test case to study the optical polarisation properties of pulsars and of their PWNe. In this paper, we review the polarisation properties of the Crab pulsar and of its 	PWN in the optical and ultraviolet domains, we summarise the state of the art of the polarisation observations of other INSs, and we outline perspectives for INS polarisation studies with present and future generations of optical telescopes. }

\FullConference{Polarimetry days in Rome: Crab status, theory and prospects\\
		 October 16-17 2008\\
		 Rome, Italy}

\begin{document}

\section{Introduction}

 Being the first and the brightest ($V \sim  16.5$) optical pulsar identified so  far \cite{coc69, mig05},  the first optical  polarisation measurements of an INS  were
obtained for   the  young ($\sim 1\,000$  yrs)  Crab pulsar (PSR\, B0531+21)  in the  Crab Nebula.   Strong polarisation is indeed expected when the optical emission is produced by synchrotron
radiation \cite{ps83}. In the case of the Crab pulsar, this is certified by its power-law spectrum $F \approx \nu^{-\alpha}$ \cite{sol00}. The brightness of the Crab pulsar made also possible to obtain phase-resolved polarisation measurements along the full pulsation period, not possible so far for the other, much fainter optical pulsars. For the same reason, the Crab is the only pulsar for which various, phase-resolved polarisation measurements have been regularly obtained. Moreover, it is the only case of an INS embedded in an optical PWN, together with PSR B0540$-$69 in the LMC. Thus, the Crab represents a prototype case to study the optical polarisation  properties of a  young pulsar  and  of its PWN and to provide deep insights into the highly magnetised  relativistic environment of young, rotation-powered INSs.  In this paper we review the observational status and the polarisation properties of the Crab pulsar and of its nebula at optical and ultraviolet (UV) wavelengths (section 2), while in section 3 we summarise the optical polarisation observations of other INSs, and in the last section we describe the perspectives of INS optical polarimetry observations with the present and future generation of telescopes and instruments.

\section{Polarisation studies of the Crab pulsar and Nebula}

Through the comparison with theoretical models, INSs optical polarimetry is a powerful diagnostic to determine, or constrain, fundamental parameters of the neutron stars, e.g. the magnetic field geometry and the neutron star rotation angle with respect to the plane of  the sky \cite{mig07}. Moreover,  the measurement of the phase-averaged polarisation position angle (PA) can allow to investigate pulsar/PWN magneto-dynamical interactions \cite{mig07}. 

The polarisation degree (PD) of the optical emission from the 
Crab was measured for the first time  \cite{wam69} soon after the
identification  of its optical counterpart \cite{coc69}.   After these exploratory works, new phase-resolved polarization measurements were performed \cite{coc70, kri70, jon81} and they all  showed that the PD is phase-dependant.  In particular, very accurate
phase-resolved polarisation measurements of the Crab performed a few years later \cite{smi88}
showed that  the pulsed  component of the  light curve  is only
weakly polarized, with  a $\le 1$ \% PD for both the main pulse (MP) and the inter pulse (IP), while the PD is much stronger ($\sim 20\%$)  in the bridge between the MP and the IP  and is maximum in the  off-pulse (OP) region, where it increases to  $\sim  40$ \%.   In addition,  the value of the  PA was also found to be  phase-dependant, with its maxima lagging $\sim 0.1$ in phase the MP and the IP.  
The most  recent measurements of the  phase-resolved polarisation  of the  Crab pulsar were obtained with the {\em OPTIMA} instrument \cite{aga09}, see also  Slowikowska et al (these proceedings), and have spectacularly confirmed  the dependance of the PD on the pulse phase.  Although a secular evolution in the optical polarisation properties of the Crab pulsar were hypothesyzed \cite{fer74}, so far repeated measurements did not unveil any  significant change. 
Phase-resolved, ultraviolet (UV) polarisation measurements of the Crab pulsar at $\sim 2700$ \AA\ were obtained with the {\em HST/HSP} \cite{gs96} . Since the {\em HSP} was decommissioned in 1993 after the first {\em HST} refurbishment mission, these are the the only UV polarisation measurements of the Crab pulsar obtained so far. The polarisation properties of the Crab pulsar in the UV are also phase-dependent and both the values of the PD and of the PA are quite similar to those observed in the optical domain, at least in the MP and IP region. Unfortunately, the {\em HSP} statistics was not sufficient to sample adequately the PD in the bridge between the MP and the IP and the OP region, although the data seem to indicate that, like in the optical, the polarisation is stronger away from the pulsation peaks.  Thus, the polarisation properties of the Crab pulsar seem to be wavelength-independent, at least for wavelengths shorter than 10000 \AA. Assuming that the phase-resolved polarisation properties depend both on the pulsar spectrum and on the geometry of the emission region, this finding, somehow, did not come as unexpected. Indeed,  there is no break in the optical-to-UV spectrum of the Crab pulsar and  no significant difference is observed between the optical and UV light curves. Interestingly, the {\em HSP} polarimetry observations of the Crab pulsar are, so far, those taken with the highest spatial resolution with an aperture of 0.65" only, i.e. small enough to minimize the contribution to the unpulsed component coming from the emission knot detected 0.5" southwest of the pulsar (see below).
Interestingly, none of the current magnetosphere models can satisfactorily account for the observed optical polarization properties of the Crab (see Kanbach, these proceedings). Normally, all magnetospheric models (e.g., the polar cap model, the outer gap model, and the two-pole caustic models) tend to predict a phase-averaged PD of $\sim$ 6.2\% which is at least a factor of 2 larger than that observed (see, e.g. \cite{mig07} for a quantitative account of the model comparison). The comparisons with theoretical models are further 
     complicated if the continuous polarisation component   is subtracted since this would yield to a PD of only 3.7\%.  The origin of this large discrepancy is yet to be explained. It could be due to a model problem in accounting for the microphysics of the emission process,  in particular the the contribution of various depolarisation effects,  to possible weaknesses of the complex  numerical codes run for model simulations, or to an observational bias like, e.g. an incorrect subtraction of the polarisation background.
The Crab pulsar proper motion has been measured several times with the {\em HST} \cite{car99, ng07, kap08}. The project proper motion vector is  close to the   symmetry axis of the PWN torus observed at optical and X-ray wavelengths by the {\em HST} and {\em Chandra}.  Interestingly, the phase-averaged polarisation vector of the pulsar \cite{smi88} is also substantially aligned with the PWN axis and proper motion vector. The same coincidence has been observed also for the Vela pulsar \cite{mig07}.  In the case of the Crab, the hard X-rays phase-averaged polarisation vector measured from {\em Integral} observations (Dean, these proceedings) is also aligned with the proper motion. Thus, the alignement between the phase-averaged PA of the pulsar, its proper motion, and the symmetry axis of the PWN can be an important tracer of the connection between the pulsar   magnetospheric  activity  and  its  dynamical  interactions  with  the PWN. 
%
%
%

The first imaging, phase-averaged,  polarimetry observations of the Crab Nebula were performed, more than a decade in advance to the pulsar discovery \cite{ow56} using photographic plates and showed the large-scale polarisation structure of the nebula. Studies of the Nebula polarisation on smaller scales \cite{sch79}  unveiled polarisation patterns on scale smaller than 20" in coincidence with the so called "wisps" (see \cite{hes08} for a general description of the Crab Nebula structure) and the "eastern bay", with a PD=30-60\%. New polarisation observations of the Nebula \cite{mcl83}, the first based on the CCD technology, unveiled finer structures in both the wisps and the eastern bay down to the 5" level.  Later polarisation observations \cite{hvdb90} unveiled large polarisation pattern also in the outskirts of the Nebula and discovered  much finer polarisation structures (scales of 2") close to the pulsar. This  clearly showed that, although the Nebula was strongly polarised and with clearly defined polarisation patterns, the  complexity of the  polarisation map was growing in the regions close to the pulsar. To study in more detail the polarisation map close to the pulsar higher spatial resolution polarisation observations, as achievable with the {\em HST} were in order. 
High resolution, narrow band imaging of the central regions of the Crab Nebula, performed with the {\em HST/WFPC2}  unveiled a peculiar emission knot at only 0.5" southwest of the pulsar, whose origin and nature are still unclear. Interestingly, a determination of the optical/infrared (IR) spectrum of the knot was obtained \cite{sol03} using imaging photometry observations obtained both with the {\em HST} and the {\em VLT}.  The knot spectrum is characterized by a power-law ($\alpha \sim -0.8$) which clearly indicates the non-thermal nature of the optical/IR emission. The know spectral slope is anticorrelated with that of the pulsar which has a spectral index of 0.3 and 0.11 in the IR and in the optical/UV, respectively. Because of its non-thermal spectrum, the knot is a natural target for polarimetry observations. Phase-averaged polarimetry observations of the central region of the Crab Nebula (including the knot) were indeed performed with the {\em WFPC2} \cite{wat96, gra96} through a narrow band filter. Polarimetry observations were also performed with the {\em HST/ACS} \cite{hes07}, luckily enough just a few months before the instrument failure of January 2007.  These observations showed that the knot is indeed polarized, with a phase-averaged PA along the symmetry axis of the X-ray torus,  like for the pulsar. With a significant polarisation, the knot might thus dominate the continuum polarisation component of the Crab pulsar \cite{aga08a} since it would not be resolved in the 2-3" apertures used in most phase-resolved polarimetry observations of the pulsar.  Unfortunately, the contribution of the knot to the continuum component can not be estimated using as a reference the higher resolution {\em HSP} observations (see previous section) since the measurement of the PD in the OP, IP, and bridge was affected by a high uncertainty. Interestingly, the knot was not found to vary on time scales of months to year and not to move.  Other interesting features unveiled by the {\em ACS} images \cite{hes07} are the polarisation of the wisps, whose PD does not seem to vary, and unexpectedly unpolarized emission knots  superimposed on the jet/counter jet.

\section{Polarisation studies of other neutron stars}

Given the impact on theoretical models, it would  be of paramount important to extend  polarimetry studies to other INSs. Unfortunately,  even though the  number of  INSs detected  in  the optical  band has  increased
significantly in  the past  ten years \cite{mig05}, 
the Crab pulsar  is still the only one for which both precise and repeated polarisation measurements have been obtained.  

For the other  young ($\le$ 10\,000 years) pulsars with an  optical counterpart, PSR\, B0540--69  ($V\sim22$), PSR\, B1509--58  ($V\sim 26$), and PSR\, B0833--45, the Vela pulsar ($V\sim 23.6$), only  preliminary, or uncertain  (i.e.  without  error  estimates),  phase-averaged  optical polarisation measurements have been reported so far. After the  first, unsuccessful attempts \cite{mid87}, polarisation observations of PSR\, B0540$-$69 were performed  with the {\em VLT} \cite{ws00} and a time-integrated  PD of $\approx$  5\%  was reported. However, this  measurement  was  strongly  contaminated by  the contribution  of   the  compact  ($\sim$  4") PWN  \cite{adl07}.  Indeed, apart from the  Crab pulsar,  PSR\, B0540--69 is  the only pulsar  featuring an optical PWN.  In particular,  the time-integrated  PD  of the nebula ($5.6\% \pm  1.0\% $ \cite{ch90}) is very similar to  that of the Crab Nebula ($\approx$ 6.9\% \cite{gok03}). High spatial resolution polarimetry observations of PSR\, B0540--69 and of its nebula have been recently obtained with the {\em HST} (Mignani et al., in prep.).  Interestingly, a bright knot in the PWN has been discovered by the {\em HST} \cite{adl07}, with a typical power-law spectrum with spectral index consistent with that of the nebula.  The knot appears displaced in {\em HST} images taken few years apart and it is not clear so far whether this effect  is related to an intrinsic variability of the nebula or to an expanding optical jet from the pulsar. 
For PSR\, B1509--58, a phase-averaged PD of $\sim 10\%$ was reported \cite{ws00} but this measurement, with the pulsar counterpart detected in the PSF wings of  a 4 magnitude brighter star, was inevitably polluted by the enhanced background contamination. 
For the Vela pulsar, a  revised and  more complete characterization  of the phase-averaged  optical polarisation properties (including the PA)  has been obtained \cite{mig07} by reanalysing  archival {\em VLT} observations  \cite{ws00}.  The  measured
fraction of phase-averaged PD ($9.4\% \pm 4\% $) was found  similar to  the  published value \cite{ws00}  but with  a larger error  obtained  from a  more  detailed  analysis.   Interestingly,  the measured value is, as in  the case of the Crab, much lower than the ones  predicted by different pulsar magnetosphere models.   Consistency with, e.g.  the outer gap   
model would require the intrinsic polarisation as low as $\sim 13\%$, i.e. much lower than expected \cite{mig07}.   Like for the Crab (Slowikowska et al., in prep.),   the
optical phase-averaged polarisation  PA \cite{mig07} coincides with that of the axis of  symmetry of the  X-ray arcs and jets  observed by {\em Chandra} and with the pulsar  proper motion direction \cite{car01,dod03}.   
For the middle-aged ($\sim 100,000$ years) pulsar PSR\, B0656+14 ($V=25$), phase-resolved polarization observations were performed \cite{ker03} which yielded to the measurement of the pulsar polarisation only over 30 \% of the lightcurve but found an extremely high PD (100 \%) in the IP region, like in the case of the Crab pulsar.  No other optical polarisation observations have been performed for other rotation-powered pulsars so far. 
For other INS classes, phase-averaged polarisation observations have been performed in the IR with the {\em VLT} for two magnetar candidates, the anomalous X-ray pulsars (AXPs) 1E\, 1048--5937 and XTE\, J1810--197 (Israel et al., in prep.) but in both cases only PD upper limits of $\sim 20\%$ have been obtained, which are significantly above the values measured for rotation-powered pulsars . 

\section{Future perspectives}

More polarimetry observations of the Crab, as well as of other INSs, to be performed at all wavelengths are important to complete the study of their magnetosphere properties.  This requires the use of the most advanced ground and space-based observing facilities. 

Most 10 m-class ground based telescopes, like the {\em VLT}, {\em SALT}, and the {\em GranteCan}, are equipped with instruments for optical/IR phase-averaged polarimetry. Strangely enough, however, no polarimetry observation of the Crab was ever performed with the {\em VLT}. With the {\em HST}, the {\em WFPC2} will be decommissioned during Service Mission 4 (SM4), early in 2009, while {\em ACS} is off-line since January 2007 and it will be hopefully repaired in SM4.  High space-resolution imaging polarimetry is still crucial to study pulsar embedded in PWNe, like the Crab itself and  PSR\, B0540--69,  to study PWN features, and to observe INSs which are too faint for phase-resolved polarimetry observations. On the other hand, phase-resolved polarimetry so far, is only possible through guest instruments, like {\em OPTIMA}, which are not accessible to the Community at large. Providing more facilities for phase averaged/resolved polarimetry is one of the possible challenges for the future generation of extra-large telescopes  like the {\em E-ELT}.

\end{document}